\documentclass[twocolumn,superscriptaddress,aps]{revtex4-1}
\usepackage{graphicx}
\usepackage{tabularx}
\usepackage[version=3]{mhchem} 

\usepackage{siunitx}

\usepackage[flushleft]{threeparttable}
\usepackage{booktabs,multirow}

\usepackage[breaklinks]{hyperref}
\hypersetup{
        pdfnewwindow=true,      
    colorlinks=true,       
    linkcolor=blue,          
    citecolor=blue,        
    filecolor=blue,      
    urlcolor=blue           
}

\begin{document}

\author{Michele Amato}
\email{michele.amato@u-psud.fr}
\affiliation{Universit\'e Paris-Saclay, CNRS, Laboratoire de Physique des Solides, 91405, Orsay, France}
\author{Thanayut Kaewmaraya}
\affiliation{Department of Physics, Khon Kaen University, Khon Kaen 40002, Thailand}
\affiliation{Institute of Nanomaterials Research and Innovation for Energy (IN-RIE), Research Network of NANOTEC-KKU (RNN), Khon Kaen University, Khon Kaen 40002, Thailand}
\author{Alberto Zobelli}
\affiliation{Universit\'e Paris-Saclay, CNRS, Laboratoire de Physique des Solides, 91405, Orsay, France}

\title{Extrinsic Doping in Group IV Hexagonal-Diamond Type Crystals}

\begin{abstract}
Over the last few years, group IV hexagonal-diamond type crystals have acquired great attention in semiconductor physics thanks to the appearance of novel and very effective growth methods. However, many questions remain unaddressed on their extrinsic doping capability and on how it compares to those of diamond-like structures. This point is here investigated through numerical simulations conducted in the framework of the Density Functional Theory (DFT).
The comparative analysis for group III and V dopant atoms shows that: 
	i) in diamond-type crystals the bulk sites symmetry ($T_d$) is preserved by doping   while in hexagonal crystals the impurity site moves towards a higher ($T_d$) or lower ($C_{3v}$) symmetry configuration dependently on the valence of the dopant atoms; ii) for Si and Ge, group III impurities can be more easily introduced in the hexagonal-diamond phase \textemdash~whose local $C_{3v}$ symmetry better accommodates the three-fold coordination of the impurity~\textemdash~while n-type impurities do not reveal any marked phase preference;
iii) for C, both n and p dopants are more stable in the hexagonal-diamond structure than in the the cubic one, but this tendency is much more pronounced for n-type impurities.
\end{abstract}

\maketitle

\section{Introduction}

In the last few years, the study of the hexagonal-diamond type allotropes (2H in the Ramsdell notation~\cite{RamsdellJEPM1947}) of group IV elements has attracted a wide interest due to the identification of novel synthesis routes not requiring critical temperature and pressure conditions~\cite{FadalyNATURE2020,WippermannAPR2016,VoroninJPCS2003,VohraPRL1986,PandolfiNL2018,BundyJCP1967}. Among these structures only the 2H-carbon phase, also called lonsdaleite, is a natural mineral occurring as microscopic inclusions in  impact diamonds. Interestingly, it can also be synthesized from highly oriented pyrolytic graphite by applying shear strain driven processes~\cite{WongCARBON2019}, shock compression~\cite{TurneaureSA2017,ShiellSR2016} or femtosecond laser pulses~\cite{SanoJPCS2009}. However, recent works argued that so far pure hexagonal-diamond C crystals might not have been obtained~\cite{SalzmannDRM2015,NemethNC2014} and that the observed materials most likely correspond to high stacking disorder in cubic diamonds (3C in the Ramsdell notation)~\cite{SalzmannDRM2015,MurriSR2019,NemethNC2014}. 
Instead, in the case of Si and Ge, it has been recently undeniably proved that 2H-nanowires (NWs) can be grown following different approaches~\cite{BarthCM2020} such as crystal transfer methods~\cite{FadalyNATURE2020,Hauge-NanoLett2015,HaugeNL17,RenNANOTECH2019}, strain and photo-induced transformation processes~\cite{Laetitia-NanoLett-2015,VincentNANOTECH2018,RodichkinaCEC2019} and plasma assisted vapor liquid-solid growth~\cite{TangNANOSCALE2017,HeNANOSCALE2019}. 

Beyond the obvious interest for fundamental science, these novel 2H allotropes might present original functionalities for electronic and opto-electronic applications with respect to their 3C counterparts~\cite{FadalyNATURE2020,TizeiEPJB2020,CaroffIEEESTQE11,JacobssonNATURE2016,AlgraNATURE2008}.
This guess has been confirmed by a number of theoretical and experimental works dedicated to the electronic~\cite{RodlPRM2019,Rodl-PRB-2015,KaewmarayaJPC2017}, optical~\cite{FadalyNATURE2020,CartoixaNL2017,Rodl-PRB-2015,AmatoNL2016,RodlPRM2019}, vibrational~\cite{FasolatoNL2018,HaugeNL17}, mechanical~\cite{ShiellSR2016} and superconducting~\cite{SakaiPRM2020} properties of these novel phase materials.
For instance, first-principles simulations have demonstrated  that 2H-Si, both in its bulk and NW form, presents an optical absorption in the visible spectral window which is higher than those of the 3C-Si allotrope~\cite{AmatoNL2016,Rodl-PRB-2015}. Yet, bulk 2H-Ge has been found to be a direct band gap semiconductor in contrast to 3C-Ge which presents an indirect band gap~\cite{KaewmarayaJPC2017,RodlPRM2019}. This has been recently experimentally confirmed by direct band gap light emission observations in hexagonal-diamond Ge and SiGe alloyed NWs~\cite{FadalyNATURE2020}. Furthermore, lonsdaleite possesses excellent mechanical properties with an hardness exceeding those of carbon diamond~\cite{PanPRL2009}.    

In view of applications, it is of the greatest importance to acquire an in-depth understanding of extrinsic doping in hexagonal-diamond type group IV crystals. So far only few articles have been dedicated to this topic and solely to the case of silicon. Fabbri et al.~\cite{FabbriNL2013}  have experimentally investigated the effect of B and P dopants on the growth dynamics of 2H-Si NWs (with a diameter ranging from 100 to 530 nm) demonstrating that the 2H-Si phase growth can be promoted by specific dopants. From the theoretical side, first-principles calculations conducted in the framework of the Density Functional Theory (DFT) have been used to study the structure and energetics of group III, group IV and group V impurities in 2H-Si bulk  and NWs.\cite{AmatoNL2019} These results demonstrate that p-type dopants can be more easily introduced in the 2H phase, both for bulk and NWs materials, as a consequence of the three-fold coordination environment that stabilizes trivalent impurities. On the contrary, no phase preference can be found for n-type dopants. These observations were supported by a stability model which shows the crucial role played by the atomic radius of the impurity. 

Whereas synthesis methods are progressing towards the growth of extended hexagonal-type structures~\cite{BarthCM2020}, 2H and 3C domains still coexist within a sub-micrometric scale in current samples. Furthermore, twin boundaries and more extended stacking faults in diamond-like crystals are hexagonal crystal inclusions with a thickness of only few atomic planes. As experimental evidences have been reported for dopant atoms segregation at crystal stacking faults~\cite{OhnoJAP2010,YamamotoJPN2014}, it is worth to investigate if a similar behaviour can exist at 2H/3C interfaces. The relative stability of dopants in the 2H with respect to the 3C phase can hence provide important insights to understand this issue.

Here, we present a comprehensive study on doping in  group IV diamond type and hexagonal-diamond type crystals. By applying DFT based methods, we investigate the formation energy and the local crystal symmetry distortion for group III (B, Al, Ga), group IV (Si, Ge, C) and group V (N, P, As) dopants in Ge and C crystals, both for the 2H and 3C phase. We compare the obtained trends with results previously obtained for Si~\cite{AmatoNL2019}.
Our main findings can be summarized as follows: i) in diamond-type crystals the bulk sites symmetry ($T_d$) is preserved by doping   while in hexagonal crystals the impurity site moves towards a higher ($T_d$) or lower ($C_{3v}$) symmetry configuration dependently on the valence of the dopant atoms; ii) for Si and Ge, group III impurities can be more easily introduced in the hexagonal-diamond phase \textemdash~whose local $C_{3v}$ symmetry better accommodate the three-fold coordination of the impurity~\textemdash~while n-type impurities do not reveal any marked phase preference;
iii) for C, both n and p dopants are more stable in the hexagonal-diamond structure than in the the cubic one, but this tendency is much more pronounced for n-type impurities. In view of the recent experimental progresses in the fabrication and characterization of 2H group IV materials, these findings suggest the optimal doping conditions for group IV allotropes and provide fundamental insights on the tendency for dopant segregation phenomena in heterophase structures.

\section{Methods}

We perform spin polarized density functional calculations under the local density approximation (LDA) using a numerical orbitals approach as implemented in the SIESTA code~\cite{SolerJPCM2002}. Only valence electrons have been taken into account with core electrons being replaced by norm-conserving pseudopotentials of Troullier-Martins type. An optimized double-$\zeta$ polarized basis set was used. Ground state geometries were optimized with a conjugate gradient algorithm adopting a convergence on the density matrix during the self-consistent cycle of $10^{-4}$ eV. A cutoff energy of 30 Ry and a $2\times2\times2$ uniform {\bf k}-grid were shown to provide total energy convergence for all considered defective structures. In order to confirm LDA results obtained for Ge (which predict an incorrect metallic band structure), additional calculations in the GGA+U approach have been performed by using the VASP code~\cite{HafnerJCC2008}. We set the U parameter to 0.4 eV and the J parameter to 4 eV, which have been demonstrated in previous works to correctly represent the Ge band gap~\cite{TahiniAPL2011}. 

Atomic coordinates and lattice parameters were relaxed adopting a force convergence criteria of 0.01 eV/\AA~and a stress convergence criteria of 0.1 GPa. As is known, the two considered allotropes present a very similar chemical environment (as reported in Fig.~\ref{Fig:figure1} for Ge crystals) which differs only when third nearest neighbors are considered. The 3C structure presents an ABC stacking along the $\langle 111 \rangle$ direction (with two atoms in the unit cell), while an ABAB atomic ordering characterizes the 2H structure along the [0001] direction (with four atoms in the unit cell). This chemical and structural proximity results in total energies, bond lengths and atomic densities that are quite close. Optimized structural parameters for all 2H and 3C reference bulk structures are reported in Table~\ref{tbl:lattice} together with the cohesion energies and atomic densities of the two considered phases. Reported values are in good agreement with previous DFT studies (see for instance Ref.~\citenum{Raffy-PRB-2004}) and experiments~\cite{BundyJCP1967,DushaqSR2019}. As shown for Si~\cite{KaewmarayaJPC2017}, all hexagonal-diamond structures present slightly lower cohesion energies with respect to cubic structures while no substantial difference in atomic density can be observed. 

\begin{table*}[tb]
	\sisetup{round-mode=places}
	\caption{Optimized lattice parameters, cohesion energies and atomic densities for bulk group IV crystals in the 2H and 3C phases. Values for Si have been extracted from Ref.~\citenum{AmatoNL2019}.}
	\label{tbl:lattice}
	\begin{tabular}{ccccccSSS}
		\hline
		\hline
		&	\text{$a_{3C}$(\AA)}  &  \text{$a_{2H}$(\AA)}   &  \text{$c_{2H}$(\AA)} & \text{$E_{coh}^{3C}$(eV)} & \text{$E_{coh}^{2H}$(eV)} & \text{$\rho^{3C}(g/cm^{3})$} & \text{$\rho^{2H}(g/cm^{3})$}  \\
		\hline
		C & 3.56 & 2.50  &   4.16  & 8.67 & 8.65 & 3.53 & 3.52  \\
		Si & 5.39 & 3.79 & 6.27  & 5.56 & 5.55 & 2.38 & 2.38  \\
		Ge & 5.76 & 4.06   &  6.69 & 4.15  & 4.13 & 5.04 & 5.06  \\
		\hline
		\hline
	\end{tabular}
\end{table*}

All the dopants were considered in substitutional sites of the host lattice. A careful analysis of formation energies convergence with respect to the cell size suggested to model the host bulk crystal with a $4 \times 4 \times 4$ supercell (containing 512 atoms) for the 3C phase  and with a $6 \times 6 \times 3$ supercell (containing 432 atoms) for the 2H phase. Both these choices, ensure the convergence of total energy, forces and local dopant symmetry and minimize spurious interaction between periodic replicas of impurities. Moreover, the size chosen corresponds to the models used in previous theoretical works on doped 3C-systems~\cite{ELPRB2004,IshiiJJAP2015,PuskaCMP2000,ProbertPRB2003}. These supercell sizes result in dopant concentration in the $10^{19}$-$10^{20}$ cm$^{-3}$ range thus they simulate the behavior of dopants in the high-doping regime observed in experiments (see for instance Refs.~\citenum{NahAPL2009,WangNL2005,NahAPL2008}).   

The standard way of predicting defects equilibrium concentrations in material consists in computing the formation energy ($E_{form}$) of the defect which,
in the case of doping, is the energy associated with the exchange of atoms between the host and a dopant atoms reservoir. Within the DFT framework, the $E_{form}$ of a neutral substitutional defect is usually formulated through the Zhang and Nortrup formalism~\cite{ZhangPRL91}: 

\begin{equation}
E_{form} = E^{D}_{tot} - E^{0}_{tot} + \mu_0 - \mu_D
\label{eq:form_en}
\end{equation}
where $E^{D}_{tot}$ and $E^{0}_{tot}$ are the total energies of a given supercell with and without the defect, and $\mu_0$ and $\mu_D$ are the chemical potentials of the host lattice atomic species and dopant, respectively. For all defects considered in this work, $\mu_0$ have been chosen as the bulk ground state total energy per atom and $\mu_D$ as the energy of the free dopant atoms. The relative stability of a given impurity between the hexagonal-diamond and cubic phases is expressed as: 

\begin{equation}
    \Delta E_{form}^{2H-3C} = E^{2H}_{form} - E^{3C}_{form} 
\label{eq:delta_form_en}
\end{equation}

This formulation has the advantage of erasing the dopant chemical potential contribution which is a parameter strongly dependent from the synthesis conditions. 
A negative value of $\Delta E_{form}^{2H-3C}$ indicates a higher stability of an impurity in the 2H phase rather than in the 3C phase, while a positive sign indicates the opposite behavior. The equilibrium concentration $[C_{i}]$ of a defect $D_{i}$ can be estimated using an Arrhenius-type relation~\cite{ZhangPRL91,ZhangPRB93,VandeWallePRB94}: 
\begin{equation}
[C_i] = N\text{exp}\left( -\frac{E_{form}(D_i)}{k_B T} \right) 
\label{eq:conc}
\end{equation}

\noindent where $N$ is the atomic density of the host lattice, $E_{form}(D_i)$ is the formation energy of the defect $D_{i}$, $k_B$ is the Boltzmann constant and T is the temperature. The relative dopants equilibrium concentration between the 2H and 3C phases can then be expressed as: 

\begin{equation}
\frac{[C^{2H}_{i}]}{[C^{3C}_{i}]} = \text{exp}\left( -\frac{\Delta E_{form}^{2H-3C}}{k_B T} \right)
\label{eq:delta_conc}
\end{equation}

\noindent This quantity  is of a particular interest since several experimental works have demonstrated the coexistence of 2H and 3C domains within individual NWs~\cite{VincentNANOTECH2018,Laetitia-NanoLett-2015,TangNANOSCALE2017,SalzmannDRM2015,MurriSR2019,NemethNC2014}. 

\section{Results and Discussion}

Results obtained for all host--dopant configurations calculated have been summarized in Table ~\ref{Table:FE_bulk_all}. Formation energies evaluated accordingly to Eq. \ref{eq:form_en} are all negative except for several dopants in carbon which indicates the difficulty in introducing such atomic species as substitutional impurities both in diamond and londsaleite. The absolute values of the formation energies for Si dopants in Ge crystals are very close to the cohesion energies of the Si lattice (Table \ref{tbl:lattice}); the use of the dopant bulk state as a reference for the chemical potential of the dopant atoms would  therefore give very low formation energies. A similar result is obtained for Ge dopants in Si. This behavior corresponds to the well known easy formation of SiGe alloys in both the cubic and hexagonal phases~\cite{FadalyNATURE2020,AmatoCR2014,HaugeNL17}.

\begin{table*} [tb!]
	\sisetup{round-mode=places}
	\caption{Calculated formation energies ($E_{form}$) (fourth and fifth columns) for group III (B, Al, Ga), IV (C, Ge) and V (N, P, As) impurities in bulk 2H and 3C-crystals of C (first block), Si (second block) and Ge (third block). The sixth column contains the difference of $E_{form}$ between the 2H and 3C phase, while the last column presents the defect concentration ratio of the two phases as derived from Eq.~\ref{eq:delta_conc}.} 
	\centering
	\begin{threeparttable}
		\renewcommand{\TPTminimum}{\linewidth}
		\makebox[\linewidth]{
		\begin{tabular}{|c|ccSSSS[table-number-alignment =center, table-figures-exponent=3]|} 
			\hline
			Host &		Dopant Group         &  Impurity & \text{$\text{E}^{2H}_{form}$ (eV)}  & \text{$\text{E}^{3C}_{form}$ (eV)} & \text{$\Delta \text{E}^{2H-3C}_{form}$ (eV)} & $[C^{2H}_{i}]/[C^{3C}_{i}]$\\ 
			\hline
			\multirow{9}{*}{  \textbf{C}} &		& B &     -6.330     &      -6.010            & -0.320  &  2.58e5 \\      
			&		\textbf{III} &   Al                  &   3.345      &     3.638      &  -0.292  &  8.71e4 \\
			&		&  Ga                   &    7.901   &        7.636       &   -0.265    &   2.96e4     \\
			\cline{2-7}
			&		\multirow{2}{*}{  \textbf{IV}}              & Si &    -1.661   &     -1.618        &  -0.043     &  5.29e0 \\       
			&		&  Ge                  &    2.841     &    2.899   &     -0.059    &  9.78e0  \\
			\cline{2-7}
			&	\multirow{3}{*}{  \textbf{V}}	& N &    -1.552     &      -1.417\tnote{a}            &   -0.135 & 1.90e2 \\    
			&		 &   P                  &   0.571     &   1.107        &  -0.535  & 1.11e9 \\
			&		&  As                   &    6.874     &  7.500    & -0.626  & 3.79e10  \\
			\hline
			\multirow{8}{*}{  \textbf{Si}\tnote{b}} &  	\multirow{3}{*}{  \textbf{III}}  & B & -6.798 & -6.530 & -0.268 &  3.44e5  \\
			&  & Al & -3.016 & -2.773 & -0.243 &  1.32e5 \\
			&  & Ga & -2.124 & -1.880 & -0.244 &  1.32e5 \\
			\cline{2-7}
			& \multirow{2}{*}{  \textbf{IV}} & C & -7.471 & -7.447 & -0.024 &  2.46 \\
			& & Ge & -3.988 & -3.992 & 0.004 & 0.89  \\
			\cline{2-7}
			& \multirow{3}{*}{  \textbf{V}} & N  & -4.358 & -4.256 & -0.102 &  5.39e0 \\
			&  & P  & -5.316 & -5.356 & 0.040 &  0.20e0 \\
			&  & As  & -2.742 & -2.805 & 0.063 &  0.08e0\\
			\hline	
			\multirow{8}{*}{  \textbf{Ge}} &	\multirow{4}{*}{\textbf{III}}	& B &     -6.733     &            -6.540     &  -0.193\tnote{c} &   1.82e3 \\   
			&		 &   Al                  &    -4.013      &       -3.846    &   -0.166 &  6.48e2     \\
			&		&  Ga                   &   -3.131    &            -2.979     &     -0.152  &  3.73e2  \\
			\cline{2-7}
			&		\multirow{2}{*}{  \textbf{IV}}     	&  C                    &    -6.356     &    -6.284   &      -0.072    &  1.64e1     \\
			&     & Si &   -5.413     &        -5.409     &   -0.003    &  1.14e0  \\       	
			\cline{2-7}
			&	\multirow{4}{*}{  \textbf{V}}	& N &     -3.797    &               -3.697   &   -0.100 &  4.85e1  \\    
			&	 &   P                 &    -5.358      &        -5.506    &  0.148\tnote{c}     & 3.12e-3  \\
			&		&  As                   &    -3.128     &          -3.296      &  0.169 &  1.41e-3   \\  
			\hline
		\end{tabular}}
		\begin{tablenotes}
			\item[a] For a seek of homogeneity with other doping configuration, this value corresponds to the N impurity forced at the center of the tetragonal site and not to its out of site ground state which is 0.67 eV lower.
			\item[b] Values extracted from Ref.~\citenum{AmatoNL2019}.
			\item[c] Using the GGA+U approximation the $\Delta \text{E}^{2H-3C}_{form}$ value is -0.17 eV for B doping and 0.09 eV for P doping.
		\end{tablenotes}
	\end{threeparttable}
	\label{Table:FE_bulk_all}
\end{table*}

All group III dopants in C, Si and Ge present negative value for $\Delta E_{form}^{2H-3C}$ implying a marked preference for the hexagonal-diamond phases. To this corresponds a very high dopant concentration ratio between the two phases (last column of Table \ref{Table:FE_bulk_all}) indicating a strong tendency for dopant segregation in the case of phase coexistence.
Group IV impurities present low absolute values of $\Delta E_{form}^{2H-3C}$ and therefore practically no doping inclination between the two phases exists.

The case of group V impurities requires a more extended discussion since no general trend emerges. Apart from the nitrogen exception, P and As in C prefer the 2H phase, in Si they have no preference of phase  and  in Ge they show a slight predilection for the 3C phase. Indeed, zero or low positive values of $\Delta E_{form}^{2H-3C}$ are obtained for P and As impurities in Si and Ge while these values are negative in C. Although there are no experimental works on the doping of the 2H-C, these findings suggest that it should be easier to introduce n-type dopants in hexagonal-diamond C crystals rather than in cubic diamonds. This result is particularly appealing because of the well known difficulty for diamond to be doped with n-type impurities~\cite{GossPSSB2008}; a similar trend was also identified for theoretically proposed wide-band gap carbon allotropes presenting a local planar bonding configuration~\cite{JungnickelPRB1998}. Nevertheless, a word of caution is in order here since $E_{form}$ values for P and As are certainly lower in the 2H phase than in the 3C phase but they still remain large and positive.

In the case of N doping, $\Delta E_{form}^{2H-3C}$ is negative for all the three host crystals which indicates once more an easier doping of the 2H phase. However, this result should be discussed in view of the possible configurations of the defective system.
It is experimentally known that all the here considered substitutional dopants in group IV 3C-type crystals preserve the $T_d$ symmetry of the host site. N doping is an exception since the symmetry is lowered to trigonal, as demonstrated by electron paramagnetic resonance (EPR) measurements~\cite{BrowerPRB1982}. From a theoretical perspective, the energy difference and relative stability between the two symmetry configurations can notably change dependently on the DFT approach employed, a point which has been largely discussed in the past (see for instance Refs.~\citenum{LombardiJPCM2003,JonesDDD2004,BacheletPRB1981,JacksonMRS1989,KajiharaPRL1991}). 
In our local basis approach the symmetry at the host site is lowered only in the case of N doped 3C-C while the $T_d$ symmetry is preserved for Si and Ge. For a seek of homogeneity with the other structures and in view of the later discussion on the influence of deformations on the stability of dopants in the two phases, the \text{$\text{E}^{3C}_{form}$} value reported in Table \ref{Table:FE_bulk_all} for carbon refers to the metastable undistorted state. The ground state formation energy of the off site system is 0.67 eV lower (this inverses the relative stability of N dopants between the 2H and 3C phases).
A final note should be addressed on the simulations of doping in Ge crystals for which the local density approximation is not able to reproduce the effective small band gap. To confirm our LDA results, we have performed additional calculations for B and P doping using the GGA+U approach which provides a better representation of the Ge electronic structure. Interesting, only minor shifts of few tenths of meV are observed for $\Delta \text{E}^{2H-3C}_{form}$ (see also the footnote c in Table~\ref{Table:FE_bulk_all}) confirming the trends discussed above.

\begin{figure*}[tb]
	\centering
	\includegraphics[width=0.8\textwidth]{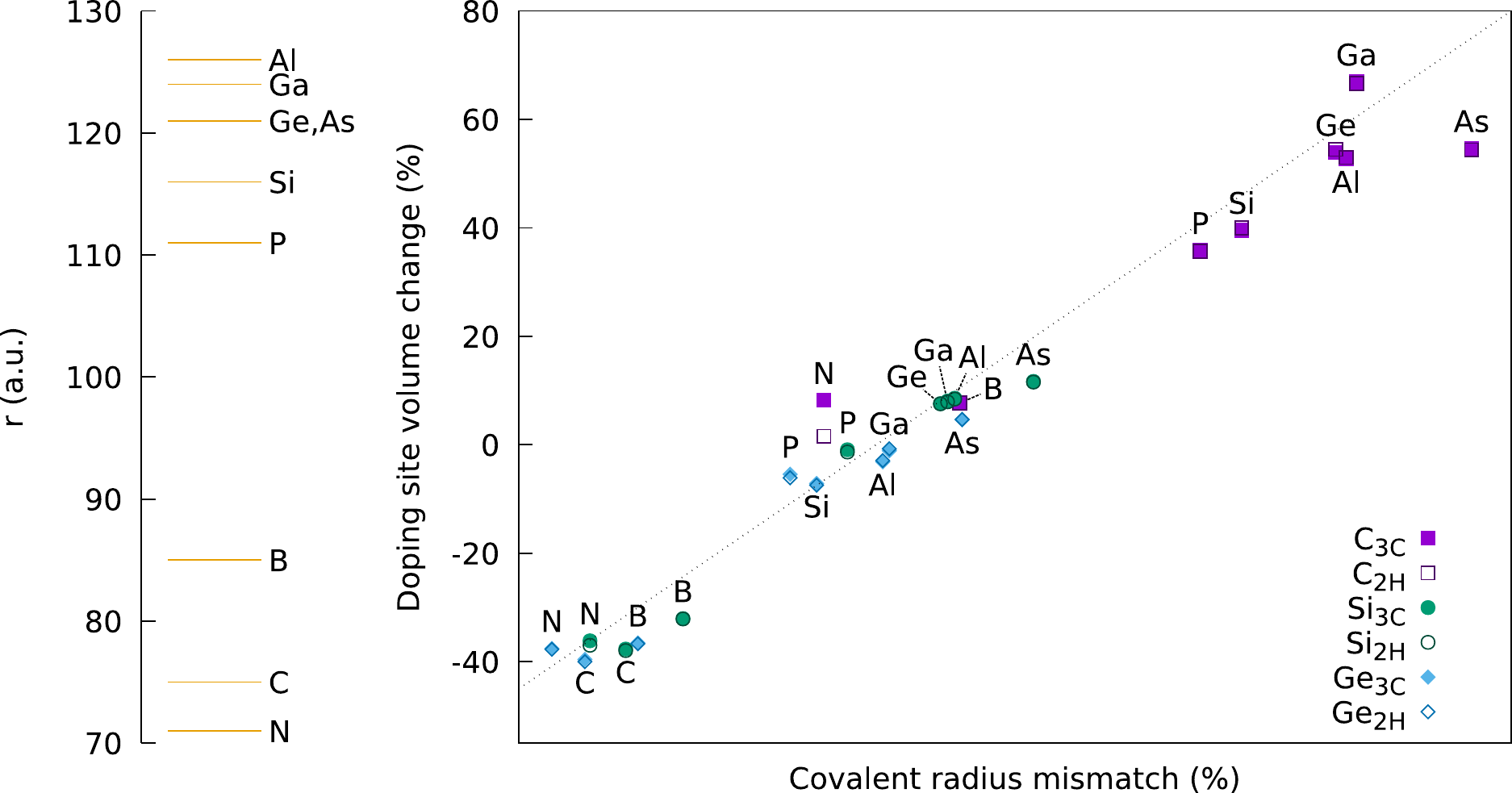}
	\caption{(Left) Covalent radius expressed in atomic units of the atomic species considered. (Right)	Atomic volume variation of the dopant site for the 3C (empty symbols) and the 2H (solid symbols) phases as a function of the covalent radius relative variation ($\Delta{r}/r$) between dopant and host atomic species. The dashed line is provided as a guide for eyes to underline the correlation between the two quantities. Except for N at C where full and empty squares are separated, in all other cases the two symbols overlap.}
	\label{Fig:Volume-change}  
\end{figure*}

The differences discussed above between the two allotropes suggest that $\text{E}_{form}$ of a defect in a given phase is an unknown function of three variables: the number of valence electrons of the impurity, the difference between the impurity atomic radius and the host atom radius (covalent radius mismatch, see left panel of Fig.~\ref{Fig:Volume-change} for a reference to atomic radius values) and the local symmetry of the impurity in the crystal.

Whereas the dependence of $E_{form}$ from these structural and electronic parameters is complex, a first insight into the different behavior in 2H and 3C crystals can be obtained by looking at the structural deformations induced by the host lattices on the dopant atoms. In the right panel of Fig.~\ref{Fig:Volume-change} we plot the covalent radius of the atomic species considered and the variation of the volume of the tetrahedron hosting the dopant as a function of the covalent radius mismatch  between the dopant and the host ($\Delta{r}$). The figure shows a clear relationship between the size of the dopant atoms and the local lattice distortion but the values relative to the 2H (full symbols) and 3C (empty symbols) phases almost perfectly overlap, except for N where they are separated. Therefore, volume changes cannot account for the $\Delta \text{E}^{2H-3C}_{form}$ values obtained.

\subsection{Symmetry Versus Valence Effects}
\begin{figure*}[tb]
	\includegraphics[width=0.8\textwidth]{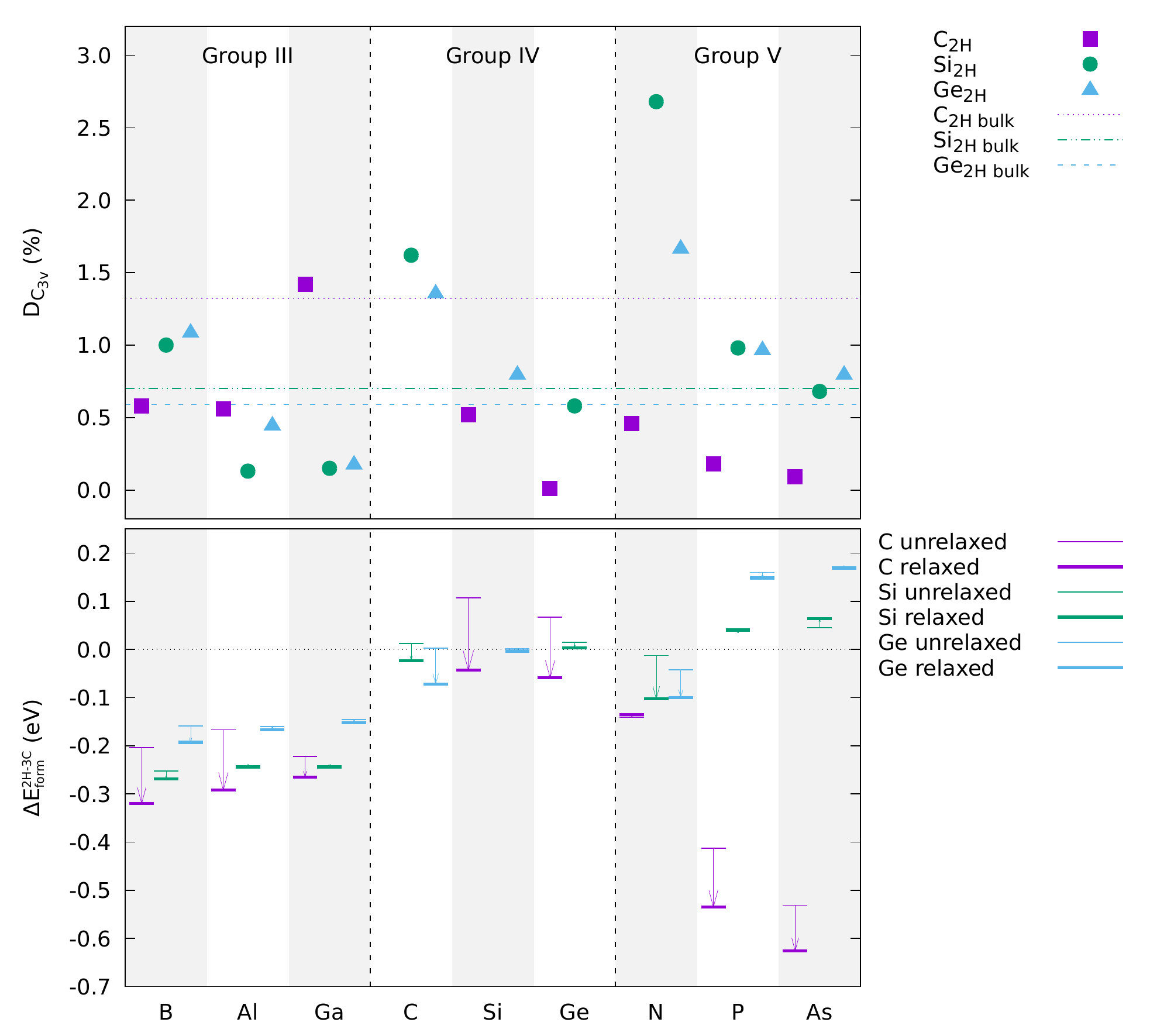}
	\caption{(Top panel) $D_{C_{3v}}$ (expressed in percent) as derived from Eq.~\ref{eq:dc3v} for different dopant atoms in 2H host structures. The horizontal lines correspond to the reference values of perfect 2H host lattices. (Bottom panel) $\Delta E_{form}^{2H-3C}$ obtained for the relaxed ground state doped structures (thick lines) and the same structure unrelaxed and constrained to the ideal bulk configuration (thin lines). Results for Si are extracted from Ref.~\citenum{AmatoNL2019}.}
	\label{Fig:DeltaE}      
\end{figure*}

\begin{figure*}[tb]
	\centering
	\includegraphics[width=0.7\textwidth]{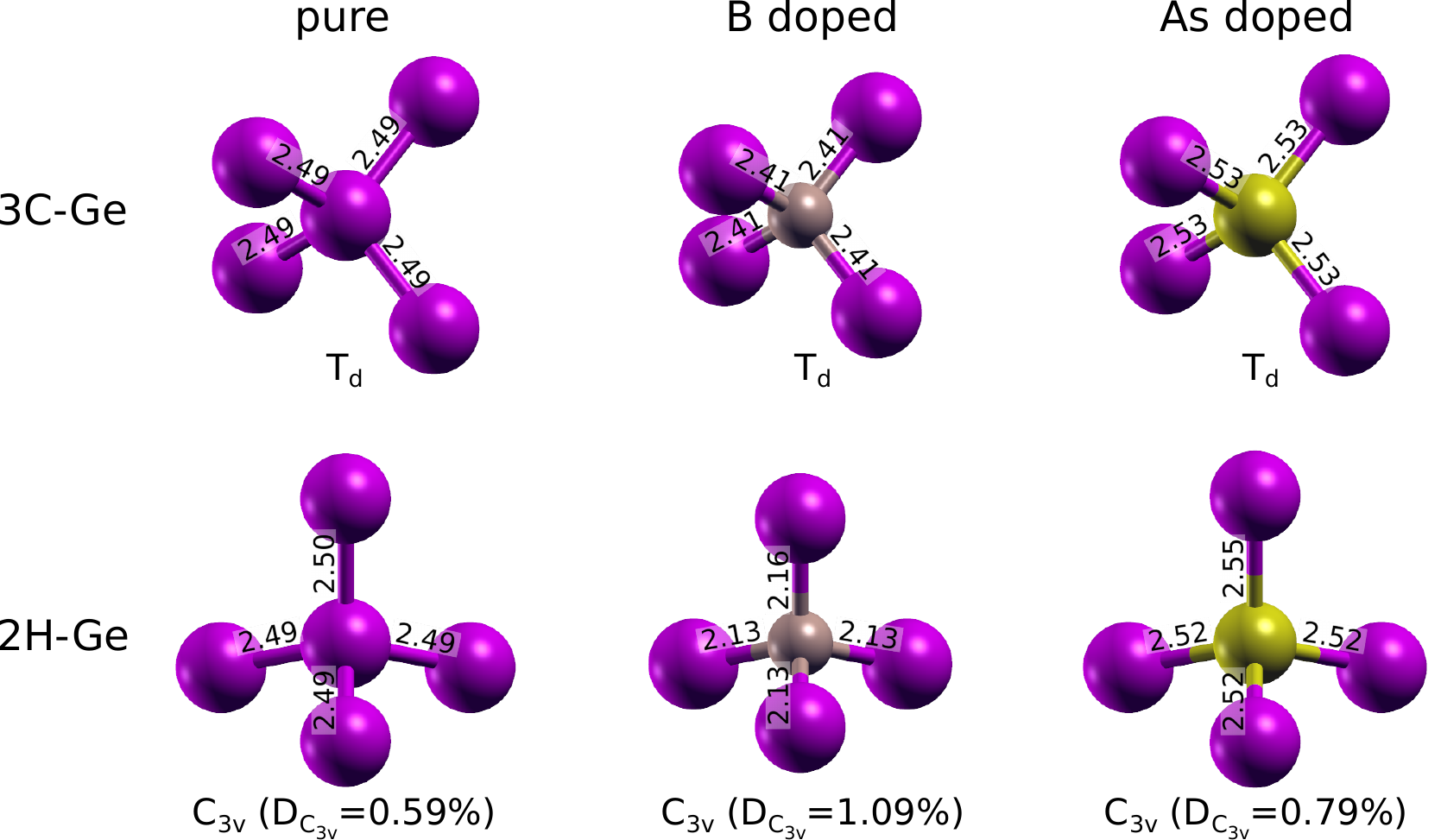}
	\caption{Tetrahedral unit of pure, B-doped and As-doped 3C-Ge (top row) and 2H-Ge (bottom row). Magenta spheres represent Ge atoms, brown spheres represent B atoms while yellow spheres represent As atoms. For each structure the corresponding symmetry as well as bond lengths (in~\AA) are highlighted.}
	\label{Fig:figure1}  
\end{figure*}

More insights can be gathered from an analysis of changes in the dopant site local symmetry. Since we are dealing with bulk systems (that are also reasonable approximation for large diameter nanowires) it is expected that the host lattice symmetry is enforced to the impurity atom site. In the 3C phase all atoms reside at the center of a perfect tetrahedron with all four first neighbors distances equal ($T_{d}$ symmetry). As reported in the first row of Fig.~\ref{Fig:figure1}, this local symmetry is preserved in the case of doping with deformations consisting in the sole uniform contraction or expansion of the bond lengths (except for N whose behavior has been discussed before). In the 2H phase, each atom has three equidistant first neighbors while the fourth neighbor, along the $c$ direction, is more apart ($C_{3v}$ symmetry, see bottom row of Fig.~\ref{Fig:figure1}). The introduction of a dopant occurs with a modification of the aspect ratio of the pyramidal site it occupies. A synthetic parameter which quantifies the deformation can be defined as follows:

\begin{equation}
     D_{C_{3v}} = {\frac{d_{\perp}-\overline{d_{\parallel}}} {\overline{d_{\parallel}}}}
\label{eq:dc3v}
\end{equation}

here $d_{\perp}$ is the longer out-of-plane bond and $\overline{d_{\parallel}}$ is the average of the three in-plane bonds at the dopant pyramidal site. In Fig.~\ref{Fig:DeltaE} (top panel) we represent the $D_{C_{3v}}$ values of 2H crystals as a function of the dopant and host species.  This quantity is larger than zero in 2H perfect crystals ($d_{\perp}$ $>$ $\overline{d_{\parallel}}$) and these reference values are indicated in the figure by horizontal lines. In a perfect 3C lattice $D_{C_{3v}}=0$ since all bond lengths are equivalent.

We start by focusing on the case of group III atom doping. These trivalent impurities naturally tend to conserve a 3-fold coordination but the bulk host lattice acts as a cage constraining the impurity site. Both structural distortion and electronic orbital rearrangements play therefore a role in defining the equilibrium configuration.
In the case of B doping we observe for instance a positive value of $D_{C_{3v}}$ which in Ge and Si host lattices is even higher than the respective reference bulk values (as a consequence of the larger radius mismatch of B with Si and Ge than with C, see left panel of Fig.~\ref{Fig:Volume-change}). Whereas the preferential $C_{3v}$ symmetry is kept for doping 2H crystals (see bottom row of Fig.~\ref{Fig:figure1} for the case of Ge), in 3C crystals only a uniform modification of bond lengths occurs and the impurity atom accommodates to the original $T_{d}$ symmetry of the host site (top row of Fig.~\ref{Fig:figure1}). The same general effect on both the 2H and 3C phases is found for the other group III dopants (Al, Ga) but its manifestation depends on the radius mismatch between the dopant and the host atom. For the same host element, lower radius mismatches between the impurity and the  host crystal atomic species lead to lower $D_{C_{3v}}$ values; in the 2H phase the dopant sites tend therefore to a more symmetric $T_{d}$ configuration.

To a first approximation, the structural deformations here described can be linked to the differences of doping formation energies  between the two phases ($\Delta \text{E}^{2H-3C}_{form}$, bottom panel of Fig. \ref{Fig:DeltaE}).
For a given group III impurity, $\Delta \text{E}^{2H-3C}_{form}$ is lower (more negative) when the bulk host lattice has a higher value of $D_{C_{3v}}$ (C, Si and Ge listing in order of magnitude). Since all values are negative, group III impurities are more stable in 2H than in 3C lattices (a word of caution is necessary for Al and Ga in C hosts as they have very large $\text{E}_{form}$ and they are hence not energetically stable). This behavior is in accordance with their trivalence which promotes the 3-fold coordination. Indeed, a higher value of $D_{C_{3v}}$ allows a better holding of the electron deficiency and an optimal adjustment of the vacant $p$ orbital. On the contrary, lower values of $D_{C_{3v}}$ force these  trivalent impurities towards the $T_d$ symmetry and therefore to form four equivalent bonds as in the case of 3C crystals.   

The analysis of symmetry effects for n-type group V dopants can be addressed using similar arguments whereas the link between structural deformation and  $\Delta \text{E}^{2H-3C}_{form}$ is more complex and no general trend emerges. 
Also for this class of impurities the symmetry of the defective site is defined by the host lattice: the $D_{C_{3v}}$ values are different from zero in 2H-crystals while they are always zero in 3C-crystals whereas, as discussed before, N doping represents an exception. Once more, the magnitude of $D_{C_{3v}}$ is related to the radius mismatch between the dopant and the host atomic species: a decrease of $D_{C_{3v}}$ occurs at the increasing of the dopant atomic radius. Notably, all $D_{C_{3v}}$ values obtained for Si and Ge are higher then the bulk references. This steric effect can be associated also to the volume expansions--contractions shown in Fig.~\ref{Fig:Volume-change}. 

The $\Delta \text{E}^{2H-3C}_{form}$ values of group V impurities in Si and Ge are larger (more positive) than those obtained for group III impurities and notably they are positive for P and As indicating a preference for the cubic phase. The chemical origin of this behavior lies in the natural tendency of group V atoms to occupy sites with a $T_{d}$ symmetry. In fact the additional electron brought by the dopant lies in an impurity level close to the conduction band and therefore it has only a minimal effect on the bonding states responsible for the local symmetry of the 3C structure. Therefore, while group III impurities accommodate preferentially in the 2H phase which allows high values of $D_{C_{3v}}$, group V dopants prefer the 3C phase which conserves the local $T_{d}$ symmetry.

Contrary to what is observed for Si and Ge crystals, in the case of C, $\Delta \text{E}^{2H-3C}_{form}$ is always negative independently from the type of dopant. The preference for the 2H with respect to the 3C  phase seems to be more pronounced for donors than for acceptor impurities: the equilibrium defect concentration ratio is of the order of $10^{9}-10^{10}$ for P and As, to be compared with $10^{5}$  for B, Al and Ga. Interestingly, the calculated values of $D_{C_{3v}}$ for P and As in 2H-C are very low revealing a clear tendency for these atoms to keep the $T_{d}$ symmetry even in hexagonal crystals. 
A reason for the discrepancy with respect to the Si and Ge cases can be identified in the large radius mismatch existing between P, As and C (left panel of Fig.~\ref{Fig:Volume-change}): at similar local symmetry conditions (equal to $T_{d}$ in the 3C phase and close to it on the 2H) a pentavalent impurity will prefer to occupy a 2H site as it allows a more extended distortion and compensation of the large radius mismatch with C. The analysis of the group V impurities in C hosts  can be indeed considered as a simple \textit{mismatch} problem. As P and As atoms have larger radii than C, they have more freedom to move in the lower symmetry structure of the 2H phase. When the group V atom becomes smaller (as in the case of N), the phase preference is notably reduced because of a considerably smaller radius mismatch. However, as pointed out before, P and As impurities present very high values of  $\text{E}_{form}$ indicating an intrinsic difficulty to dope diamond and hexagonal-diamond crystals. 

\subsection{Relaxation Effects}
Finally, further insights on the parameters defining the relative dopant stability between the phases can be gathered by comparing the values of $\Delta E_{form}^{2H-3C}$  obtained for the relaxed ground state with those of the same structure constrained to the ideal bulk configuration (bottom panel of Fig.~\ref{Fig:DeltaE}). This analysis has the aim of separating the contribution of geometry and electronics in the energetic cost of creating the defect. Interestingly, in the case of Ge (light blue lines) the effect of atomic relaxation is very pronounced for the lightest impurities (B, C and N), while it can be considered negligible in all other cases. The behavior is very similar to the trend observed in Si (green lines in Fig.~\ref{Fig:DeltaE}, see also Ref.\citenum{AmatoNL2019}).
This means that the electronic effect, which is related to the dopant valence, can be the driving force in defining the preferential phase for doping, except in the case of large differences between the dopant and host atomic radius. 
As a further proof of it, the trend observed for C clearly shows that the unexpected stability of group V impurities in the 2H phase is a consequence of the large radius mismatch and the consequent geometrical relaxation effects. As discussed above, the relaxation can be very relevant in this case because a large impurity tends to deform the host and in that case it is easier to do it in the 2H structure.  

\section{Conclusions}
In this work, we have presented a comprehensive study, conducted in the framework of initio DFT calculations, on the doping of Ge and C hexagonal-diamond type crystals and we have compared these results with those previously obtained for Si crystals~\cite{AmatoNL2019}. Our simulations reveal that  hexagonal-diamond polymorphs of Si, Ge and C can be ideal hosts for p-type dopants. Instead, group IV impurities do not present any phase preference: substitutional alloys can form both in the cubic and hexagonal-diamonds lattices as a result of very close chemical potentials, less than 0.02 eV of difference as derived from bulk references. Concerning n-type dopants, their associated formation energies do not show clear phase preference in the case of Si and Ge crystals while for C a higher dopability is found in the 2H phase. A careful structural analysis reveals that, for all dopant species considered, volume changes at the doping site are almost equal in the two phases and therefore this quantity cannot account for the formation energy trends observed. 

The discussed results are devoted to the behavior of impurities in bulk host crystals. However, since the size of NWs that are routinely grown is in the 20-100 nm range, this model represents also a practical and effective mean to describe doping mechanisms in middle- and large-diameter NWs where quantum confinement and surface effects are negligible. As it has been noticed in recent theoretical works~\cite{ZhaoPRB2019,AmatoNL2019}, when surface effects or significant distortions are considered, dopant sites become nonequivalent making the role of the crystal phase symmetry secondary with respect to volume effects. This is particularly true in the case of other nanostructured systems as Si and CdSe nanocrystals~\cite{CantelePRB2005,DalpianNL2006}. In our case, the local symmetry at the impurity site is a more pertinent quantity to understand the relative stability of dopants in the two phases: in diamond-type crystals the bulk sites symmetry ($T_d$) is preserved by doping   while in hexagonal crystals the impurity site moves towards a higher ($T_d$) or lower ($C_{3v}$) symmetry configuration dependently on the valence of the dopant atoms. Therefore, if we disregard the case of group IV impurities, the lower symmetry of the 2H phase is almost always preferred as it represents the ideal environment to satisfy both electron valence and structural relaxation requirements.

All calculations and models here presented shed light on the doping properties of hexagonal-diamond crystals and they suggest the optimal doping conditions that characterize these materials. The described effects could be at the origin of possible segregation phenomena at 2H/3C nanowire interfaces for which further theoretical and experimental investigations are needed. 

\acknowledgements
We are grateful to E. Canadell, R. Rurali and S. Ossicini for their critical reading of the manuscript and for helpful discussions. We greatly acknowledge the ANR HEXSIGE project (ANR-17-CE030-0014-01) of the French Agence Nationale de la Recherche. Part of the high-performance computing (HPC) resources for this project were granted by the Institut du developpement et des ressources en informatique scientifique (IDRIS) under the allocation A0040910089 via GENCI (Grand Equipement National de Calcul Intensif).

\bibliography{manuscript}

\end{document}